\documentclass[aps,eqsecnum,amsmath,onecolumn,notitlepage,nofootinbib]{revtex4-1}
\usepackage[utf8]{inputenc}
\usepackage{indentfirst}
\usepackage{amssymb}
\usepackage{amsmath,latexsym}
\usepackage{graphicx}
\usepackage{caption}
\usepackage{subcaption}
\usepackage{float}
\usepackage[margin=1.0in]{geometry}
\usepackage[mathscr]{euscript}
\usepackage{color}
\graphicspath{{images/}}

\newcommand{\beq}{\begin{equation}}
\newcommand{\eeq}{\end{equation}}
\newcommand{\bea}{\begin{eqnarray}}
\newcommand{\eea}{\end{eqnarray}}
\def\beqs#1\eeqs{\beq\begin{split} #1 \end{split}\eeq}

\def\Man{{\cal M}}
\def\R{\mathbb R}
\def\C{\mathbb C}

\def\Im {\mathop{\hbox{Im}}}

\def\av#1{ \left\langle #1 \right\rangle }

\usepackage[colorlinks=true,backref=false, linktocpage=true,
citecolor=blue,urlcolor=blue,linkcolor=blue,pdfpagemode=UseOutlines]{hyperref}

\hypersetup{%
  bookmarksnumbered=true,
  pdftitle = {},
  pdfsubject = {},
  pdfauthor = {},
  pdfkeywords = {}
}

\usepackage{microtype}

\newcommand{\nn}{\nonumber}

\DeclareMathOperator{\im}{Im}
\DeclareMathOperator{\re}{Re}

\begin{document}

\title{Tempered transitions between thimbles}

\author{Andrei Alexandru}
\email{aalexan@gwu.edu}
\affiliation{Department of Physics, George Washington University, 
Washington, DC 20052}
\author{G\"{o}k\c{c}e Ba\c{s}ar}
\email{gbasar@umd.edu}
\affiliation{Physics Department,University of Illinois at Chicago, Chicago, Il 60607}
\affiliation{Department of Physics, University of Maryland, College Park, MD 20742}
\author{ Paulo F. Bedaque }
\email{bedaque@umd.edu}
\author{Neill C. Warrington}
\email{ncwarrin@umd.edu}
\affiliation{Department of Physics, University of Maryland, College Park, MD 20742}


\begin{abstract}
Quantum field theories with complex actions cannot be investigated
using importance sampling due to the sign problem.
One possible solution is to use the holomorphic gradient flow, a
method we introduced related to the Lefschetz thimbles idea.
In many cases the probability distribution generated by this method
is multi-modal and standard Monte-Carlo sampling fails. We propose
an algorithm that incorporates tempered proposals to solve this problem.
We apply this algorithm to the $0+1$ dimensional Thirring model at finite density for
a parameter set where standard sampling fails and show that 
tempered proposals cure this problem.
\end{abstract}

\maketitle

\section{Introduction}

Non-perturbative results from quantum field theories can be obtained
using stochastic sampling, as long as the path integral can be represented
as a sum over real and positive contributions. This is no longer the
case when one considers phenomenologically interesting problems regarding
systems at finite density, like QCD at non-zero baryon density, or 
questions related to real-time dynamics. In these cases, the integrand 
is complex and direct Monte-Carlo sampling cannot be applied. The standard 
work-around, {\em reweighting}, that samples according to a positive measure and 
corrects for the difference by introducing a complex fluctuating phase in the 
observables, fails due to phase oscillations; this is the infamous 
{\em sign problem}. Recently a possible solution was 
proposed by Cristoforetti~et~al.~\cite{Cristoforetti:2012su}: We start
by embedding the integration domain of the path integral in a complex space.
Using the analytical properties of the integrand we then deform the integration
manifold without changing the integral value. The proposal by Cristoforetti
was to use the a manifold that corresponds to a union of Lefschetz thimbles. 
This {\em thimble decomposition} is always possible and it has the advantage 
that the integrand's complex phase
on each thimble is constant. When the integral is dominated by the contribution
of one thimble, the sign problem is effectively solved~\cite{Cristoforetti:2013wha,Mukherjee:2013aga,Fujii:2013sra,Alexandru:2015xva,Alexandru:2016san}.

In cases where more than one thimble contributes significantly to the
integral, the problem is significantly harder. Sampling algorithms have
to be able not only to sample each thimble but also be able to dynamically
transition between thimbles in order to properly take into account
their relative contribution. Additionally, significant analytical work
is needed to identify all the thimbles contributing to the integral.
In an earlier paper~\cite{Alexandru:2015sua} we proposed a method that 
sidesteps this task. The idea is to use a class of manifolds generated
by the {\em holomorphic gradient flow} and parametrized by
the flow time $T_\text{flow}$ interpolating smoothly between the
original integration domain~($T_\text{flow}=0$) and the thimble 
decomposition~($T_\text{flow}=+\infty$). While the value of the integral 
on all these manifolds is the same, the phase fluctuations become progressively
smaller as we increase $T_\text{flow}$, allowing us to use reweighting. 
On the other hand as $T_\text{flow}$ increases
the probability distribution becomes multi-modal with growing potential
barriers between different modes. For algorithms that rely on small-step updates,
such distributions are difficult to sample since
the transitions rate between modes becomes very small. For some systems there
are values of $T_\text{flow}$ for which the sign problem is mild and the
transition rate between modes is good such that standard small-step algorithms 
can be used~\cite{Alexandru:2016gsd,Alexandru:2016ejd}.

For systems where the sign problem only becomes manageable when 
$T_\text{flow}$ is large and the probability distribution is multi-modal 
a possible solution is to use the method of {\em tempered transitions}. 
The basic idea is to use a set of small-steps to build a 
large mode-to-mode move~\cite{Neal1996}.
The sequence of steps is constructed using a set of guiding distributions that
overlap well sequentially but gradually lower and raise the potential
barriers between modes. In this paper we discuss a proposal where the 
guiding probabilities are generated by changing the $T_\text{flow}$ 
and apply it to the Thirring model in $0+1$ dimensions, a system whose
path integral decomposition requires multiple 
thimbles~\cite{Alexandru:2015xva}. The plan of the paper is the following:
in Section~\ref{sec:manifolds_alone} we review the relevant details for the
holomorphic flow and thimble decomposition, in Section~\ref{sec:tempering_paulo}
we review the details of the tempered transitions algorithm as it applies 
to our problem, and in Section~\ref{sec:results} we review the relevant
details for the $0+1$ Thirring model and present the results of our simulations. 

\section{Holomorphic Gradient Flow}
\label{sec:manifolds_alone}

Here we show how to deform the domain of path integration in order to ameliorate the sign problem~\cite{Cristoforetti:2013wha,Alexandru:2015sua}. The starting point is Cauchy's theorem, which allows one to deform the domain of path integration ($\R^N$)  into a submanifold $\Man$ of complex space ($\C^N\approx \R^{2N}$) without changing the value of the path integral:
\beq
\langle \mathcal{O}\rangle =
\frac{\int_{\R^N} d\zeta_i \  e^{-S[\zeta]} \mathcal{O}[\zeta] }{\int_{\R^N}  d\zeta_i \  e^{-S[\zeta]}}
=
\frac{\int_{\Man} d\phi \  e^{-S[\phi]} \mathcal{O}[\phi] }{\int_{\Man}  d\phi_i \  e^{-S[\phi]}},
\eeq where $\zeta_i$, $i=1,\ldots, N$ are real field variables. The sign problem arises because $S[\zeta]$ is not real, leading to rapid phase oscillations in the path integral.  The goal is to find a manifold, $\Man$, such that Cauchy's theorem applies and $S[\phi]$ does not oscillate as rapidly for $\phi \in \Man$ as it does for $\zeta \in \R^N$. One way to construct such a manifold is to identify every field configuration in the original integration domain ($\R^N$) as an initial condition for the following set of first order differential equations known as the \textit{holomorphic gradient flow equations:}
\bea\label{eq:flow}
\frac{d\phi_i}{dt} = \overline{ \frac{\partial S}{\partial\phi_i}}, \quad \phi_i(0) = \zeta_i,
\eea
integrated up to a fixed ``flow time" $T_{\text{flow}}$. Here the bar on the RHS of Eq.~\ref{eq:flow} denotes complex conjugation. These equations map a particular field configuration, $\zeta\in\R^N$ to a point $\phi(T_{\text{flow}})\in\C^N$. We will call this motion the ``flow". The map defined by the flow $\zeta\mapsto\phi(T_{\text{flow}})$ is one-to-one as Eq.~\ref{eq:flow} is first order. Therefore flowing $\R^N$ generates a N real-dimensional manifold in $\Man \subset \C^N$ (i.e. an $N$ real-dimensional manifold, $\Man$, embedded in $\C^N\simeq \R^{2N}$). 

Having constructed $\Man$ we now to establish that (i) Cauchy's theorem applies on $\Man$ $($so the path integral on $\Man$ is equal to the path integral on $\R^N$$)$ and (ii) the phase oscillations on $\Man$ are milder than the phase oscillations on $\R^N$, which leads to a milder sign problem. First observe an important property of the flow equations: the real part of the action, $S_R$, increases monotonically along a  flow trajectory, whereas the imaginary part, $S_I$, stays constant:
\bea\label{eq:SR_SI}
\frac{dS_R}{dt} ={1\over2}\left( { \frac{\partial S}{\partial\phi_i}} {d\phi_i\over dt}+ \overline{ \frac{\partial S}{\partial\phi_i}} \,\overline{d\phi_i\over dt}\right)=\left|\frac{\partial S}{\partial\phi_i}\right|^2>0
\quad,\quad
\frac{dS_I}{dt} ={1\over2 i}\left( { \frac{\partial S}{\partial\phi_i}} {d\phi_i\over dt}-\overline{ \frac{\partial S}{\partial\phi_i}} \,\overline{d\phi_i\over dt}\right)=0
\eea
where in the second equalities we used Eq.~\ref{eq:flow}. For (i) to hold, it must be that points do not cross any singularities throughout the continuous deformation process, especially singularities that might arise when the field variables go to infinity. Indeed no singularities are crossed because $S_R$ increases monotonically with the flow Eq.~\ref{eq:SR_SI}, which ensures that the modulus of the integrand, $|e^{-S}|=e^{-S_R}$ decreases monotonically, and therefore remains bounded from above and damps the integral exponentially as the fields approach infinity \footnote{We assumed that the path integral was convergent to begin with. This is indeed the case when the lattice spacing is finite. The standard renormalization procedure has to be followed to approach the continuum limit.}. 

For property (ii), consider the limit of large flow time, $T_{\text{flow}}$. In this limit, almost all the original field configurations in $\R^N$ will flow into configurations with very large $S_R$ and will contribute practically nothing to the path integral due to the exponential damping factor $e^{-S_R}$. Therefore in the large flow limit, the main support of the path integral will come from fields that flow to fixed points of Eq.~\ref{eq:flow} which are critical points of the action $\partial S/\partial\phi_i=0$ (i.e. classical solutions to the complexified equations of motion).  Consider a point in $\R^N$ that flows to a critical point. Then an infinitesimal neighborhood around this point  will flow to an $N$-real dimensional manifold ${\cal J}$, attached to that critical point. Since $S_I$ remains unchanged with the flow and the variation of the $S_I$ in an infinitesimal neighborhood is infinitesimal, $S_I$ will be approximately constant on ${\cal J}$. In the limit $T_{\text{flow}}\to\infty$, $S_I$ will be exactly constant. This $N$ dimensional manifold ${\cal J}$ attached to a critical point, over which $S_I$ is constant is known as a ``Lefschetz thimble" $($the multi-dimensional generalization of the stationary phase/steepest descent contour familiar from complex analysis$)$. In the limit $T_{\text{flow}}\to\infty$, $\Man$ will be a particular combination of thimbles, and $S_I$ will be piecewise constant on $\Man$. For sufficiently large, but finite $T_{\text{flow}}$, $S_I$ will not exactly be piecewise constant, but will be approximately piecewise constant. Consequently, by tuning $T_{\text{flow}}$ we can continuously soften the severity of the phase oscillations caused by $e^{iS_I}$. In other words, the manifolds defined by different $T_{\text{flow}}$ interpolate between $\R^N$ ($T_{\text{flow}}=0$) where $S_I$ varies rapidly and the associated thimble decomposition of the path integral ($T_{\text{flow}}\to \infty$) where $S_I$ is piecewise constant.

\begin{figure}[h]\captionsetup{singlelinecheck = false, justification=raggedright}
\includegraphics[scale=0.55]{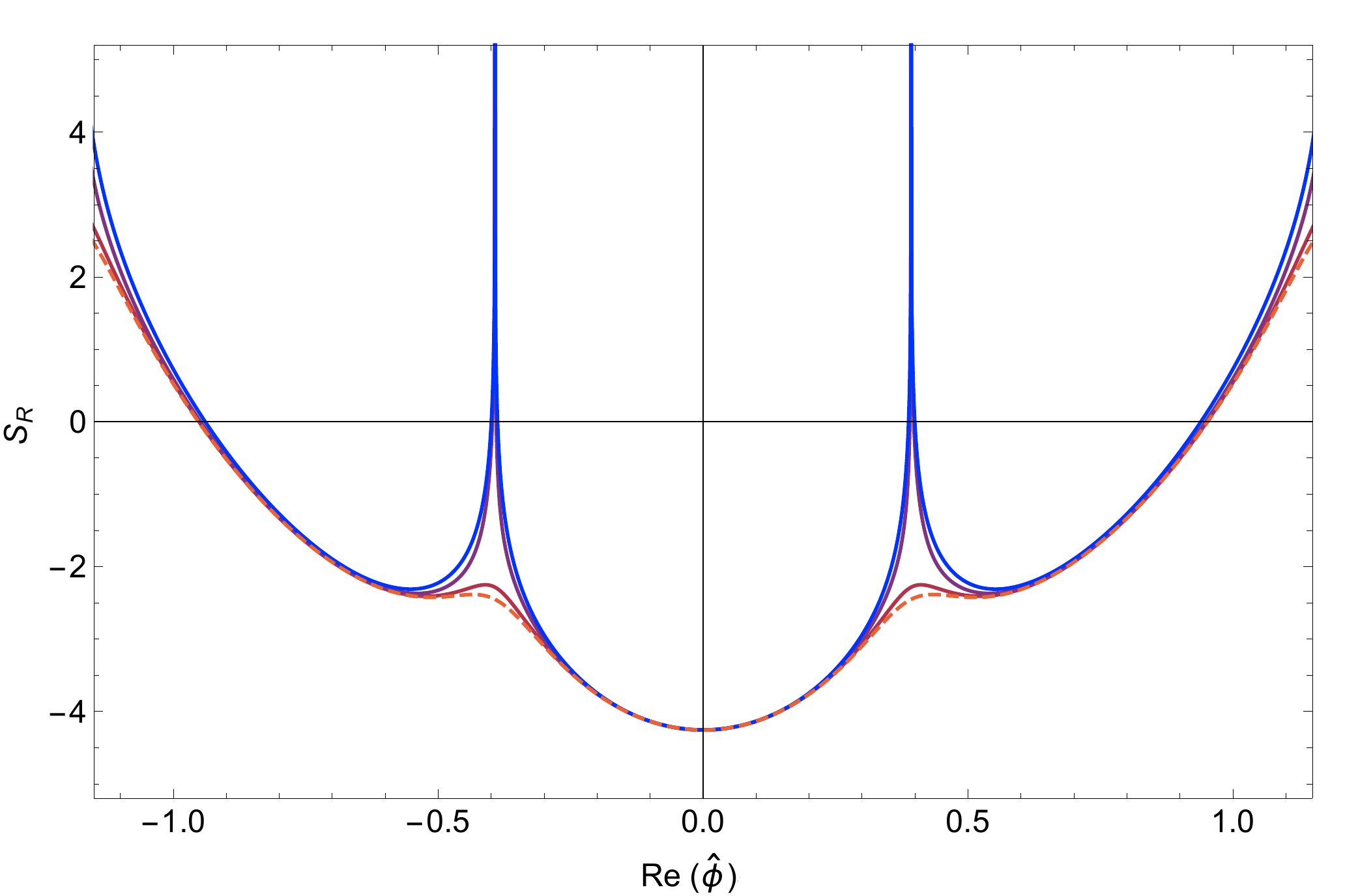}
\caption{Plot of the real part of the action on the one dimensional subspace of $\Man$ obtained by projecting onto constant fields~\cite{Alexandru:2015sua}. The red dotted line is the real part of the action along the tangent space of the thimble while the progressively bluer lines are obtained by flowing the tangent space by increasing amounts. The action barriers quickly diverge to infinity and are the reason for fields getting trapped to a single thimble.}
\label{fig:srvsphi}
\end{figure}

It is desirable, from the point of view of the sign problem, to integrate over a highly flowed manifold because the only regions with appreciable support have nearly constant $S_I$. However there are costs to this procedure. First, it is numerically expensive to do so. More fundamentally, however, regions of support on the parameterization manifold $($where the Monte Carlo takes place$)$ of a highly flowed surface are separated in field space by extended regions with very large $S_R$. Consequently a Monte Carlo simulation must sample from a multi-modal distribution. The situation is worse when there are fermions involved, because in these cases, thimbles have \textit{boundaries}: $N-1$ dimensional sub-manifolds where the fermion determinant vanishes and $S_R\rightarrow\infty$. As an illustration of this phenomenon we shown in Fig.~\ref{fig:srvsphi} the action on a submanifold of $\Man$ for various flow times. One way to avoid this situation is, instead of approaching the $T_{\text{flow}}\to\infty$ limit, to tune $T_{\text{flow}}$ to a finite value where the phase oscillations are under control yet the action barriers are not so high that the Monte Carlo gets trapped. In a variety of examples this is indeed possible \cite{Alexandru:2016gsd,Schmidt:2017gvu,Alexandru:2016san}. However, depending on the model and, in particular at large volumes, it may not be possible to solve the sign problem and the isolated minima problem simultaneously by tuning the flow time. Thus we come to the central point of this paper: how to tackle multimodal distributions created by the holomorphic gradient flow.

Before concluding this section we outline the Monte Carlo computation that we will employ later on. Using the fact that the flow defines a one-to-one mapping between the initial field, $\phi_i(0)=\zeta_i$ and the flowed field $\phi_i(T_{\text{flow}})$ we can parameterize the path integral over $\Man$ using real variables $\zeta_i$
\bea
Z = \int_{\Man} d\phi_i \ e^{-S[\phi] }
= \int_{\R^N} d\zeta_i\      {\rm det}\left ( \frac{\partial \phi_i}{\partial \zeta_j}\right )  e^{-S[\phi(\zeta)]}.
\eea 
Notice that this a re-parameterization of $\Man$ and is distinct from the contour deformation from $\R^N$ to $\Man$ that we discussed earlier.  Parameterizing $\Man$ with real fields allows us to perform the Metropolis updates on $\R^N$. The Jacobian, $J_{ij}=\partial \phi_i/ \partial \zeta_j $,  associated with this change of variables also satisfies a flow equation
\bea\label{eq:flowJ}
\frac{dJ_{ij}}{dt} = \overline{H_{ik}}\overline{ J_{kj}}, \quad H_{ij}\equiv \frac{\partial ^2S}{\partial\phi_i\partial\phi_k},\quad J_{ij}(0) = \delta_{ij},
\eea 
which transports the local tangent space at $\zeta_i$ to the flowed point $\phi_i(T_{\text{flow}})$ along the flow trajectory followed by $\zeta_i$. The determinant of this Jacobian is a complex number which we combine with the action to define an effective action, $S_\text{eff}[\zeta] =  S[\phi(\zeta)] - \log\det J$. In our Monte Carlo simulations, the configurations are sampled according to the real part of the effective action, $\re S_\text{eff}[\zeta] = S_R[\phi(\zeta)]-\log|\det J|$. The phase of the Jacobian, along with the phase of the action, $\varphi(\zeta)\equiv\im S_\text{eff}[\zeta]=  S_I[\phi(\zeta)]-\Im \det J$, is included via ``reweighing":
\bea
\av{ \mathcal{O}} &=
\frac{
\int d\zeta_i\   \mathcal{O} \det J e^{-S[\phi(\zeta)]}
}
{
\int d\zeta_i\    \det J e^{-S[\phi(\zeta)]}
}
=
\frac{
\int d\zeta_i\   \mathcal{O} e^{-i \varphi(\zeta)}  e^{-\re S_\text{eff}[\zeta]}
}
{
\int d\zeta_i\   e^{-\re S_\text{eff}[\zeta]}
}
\frac{
\int d\zeta_i\   e^{-\re S_\text{eff}[\zeta]}
}
{
\int d\zeta_i\    e^{-i \varphi(\zeta)} e^{-\re S_\text{eff}[\zeta]}
}=
\frac
{\av{ \mathcal{O} e^{-i \varphi(\zeta)} }_{\re S_\text{eff}}}
{\av{ e^{-i \varphi(\zeta)} }_{\re S_\text{eff}}}.
\eea 
Finally, even though we are performing a path integral over $\R^N$ due to the way we parameterize $\Man$, the action and all the operators are still evaluated on $\Man$. In particular the fluctuations of $S_I[\phi(\zeta)]$ which enter in are drastically milder than those over the original domain given by $S_I[\zeta]$. 

\medskip
\section{Tempered transitions}
\label{sec:tempering_paulo}

The method of tempered transitions was designed to perform Monte Carlo calculations in situations where the desired probability distribution is multimodal, that is, has more than one, well separated peaks~\cite{Neal1996}. Multimodal distributions are challenging for Monte Carlo methods because, with most algorithms, the Monte Carlo chain ends up being trapped in one of the modes for a very large number of steps making it nearly impossible to sample properly. 

We now quickly describe the method of tempered transitions in general terms, before applying it to a specific model.
 Suppose the distribution of interest is $p(\phi)$.  Then a standard importance sampling technique is to make  symmetric proposals $\phi \rightarrow \phi'$ in the sampling space and accept such proposals with probability $\text{min}\{1,\frac{p(\phi')}{p(\phi)}\}$. In order to achieve a reasonable acceptance rate, the proposed $\phi'$ is chosen close to $\phi$. For a multimodal distribution this leads to the trapping alluded above. In the tempered transitions methods one makes a more sophisticated $($and computationally expensive$)$ proposal that has a fair likelihood of being on another mode and also of being accepted. This is achieved by considering a sequence of $n+1$ progressively flatter probability distributions $p_i(\phi)$, with $p_0(\phi)=p(\phi)$ being the desired distribution to sample from. For each of these distributions $p_i(\phi)$ consider a transition probability $\hat{T}_{i+1}(\hat{\phi}_{i}\rightarrow\hat{\phi}_{i+1}) $ satisfying detailed balance with respect to $p_i(\phi)$:
 \beq
p_i(\phi_i) \hat{T}_{i+1}(\hat{\phi}_i\rightarrow\hat{\phi}_{i+1}) = p_i(\phi_{i+1}) \hat{T}_{i+1}(\hat{\phi}_{i+1}\rightarrow\hat{\phi}_i) .
\eeq The transition probabilities $\hat{T}_{i}(\hat{\phi}_{i-1}\rightarrow\hat{\phi}_i)$ can be chosen to be, for instance, Metropolis steps.

\begin{figure*}[t]
\includegraphics[scale=0.35]{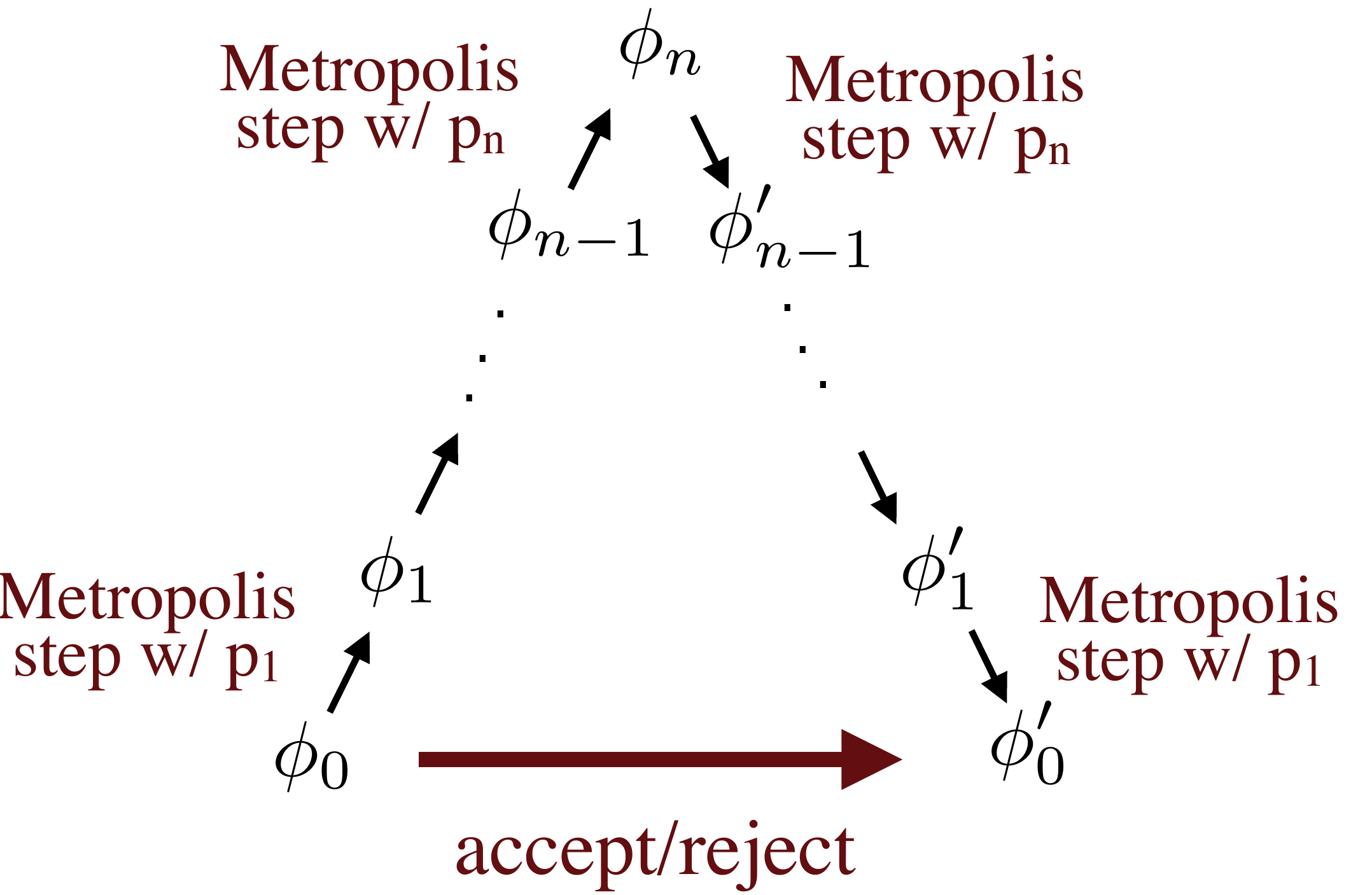}
\caption{The tempered transition Monte Carlo step. }
\label{fig:tempering_step_figure}
\end{figure*}

  The proposed configuration $\phi'$ is obtained from $\phi=\phi_0$ by evolving $\phi$ through a series of updates with the transition probabilities $\hat{T}_{i}(\hat{\phi}_i\rightarrow\hat{\phi}_{i+1})$  all the way up to $\hat{T}_{n}(\hat{\phi}_{n-1}\rightarrow\hat{\phi}_{n})$ and then again, in reverse order, down to $\hat{T}_{1}(\hat{\phi}_{1}'\rightarrow\hat{\phi}_{0}')$  (see Fig.~\ref{fig:tempering_step_figure}). More precisely, the probability of proposing $\phi'=\phi_0'$ starting from $\phi=\phi_0$ along the path $\phi_0 \rightarrow \phi_1 \rightarrow ... \rightarrow {\phi_1}' \rightarrow {\phi_0}'$ is given by:
  \bea\label{eq:tempered_proposal}
\mathcal{P}(\phi_0\rightarrow \phi_0') &=& 
                                                                        \hat{T}_{1}(\hat{\phi_0}\rightarrow\hat{\phi_1})  
                                                                        \hat{T}_{2}(\hat{\phi_1}\rightarrow\hat{\phi_2}) \cdots
                                                                         \hat{T}_{n}(\hat{\phi}_{n-1}\rightarrow\hat{\phi_n}) \nn\\
                                                                        && \qquad\qquad
                                                                        \hat{T}_{n}(\hat{\phi_{n}}\rightarrow\hat{\phi}_{n-1}) 
                                                                          \cdots
                                                                         \hat{T}_{2}(\hat{\phi_1'}\rightarrow\hat{\phi_2'}) 
                                                                         \hat{T}_{1}(\hat{\phi_1'}\rightarrow\hat{\phi_0'}) ,
\eea
The final configuration $\phi_0'$ is the proposed configuration that is accepted or not according to the acceptance probability
  \beq
\mathcal{A}(\phi_0\rightarrow \phi_0') = {\rm min}\left(
1, \frac{p_1(\phi_0)}{p_0(\phi_0)} \cdots \frac{p_n(\phi_{n-1})}{p_{n-1}(\phi_{n-1})}  \frac{p_{n-1}(\phi_{n-1}')}{p_{n}(\phi_{n-1}')} \cdots  \frac{p_0(\phi_0')}{p_1(\phi_0')} 
\right).
\eeq It is straightforward to verify that the combination of a tempered transition and the accept/reject step satisfy detailed balance and thus samples the true distribution $p(\phi)$:
\beq\label{eq:tempered_detailed}
p(\phi_0) \mathcal{P}(\phi_0\rightarrow \phi_0') \mathcal{A}(\phi_0\rightarrow \phi_0')  =
p(\phi_0') \mathcal{P}(\phi_0'\rightarrow \phi_0) \mathcal{A}(\phi_0'\rightarrow \phi_0) .
\eeq
While Eq.~\ref{eq:tempered_detailed} guarantees the correctness of the method, its usefulness relies on its ability to generate proposals transitioning between modes with a high probability of acceptance. A heuristic discussion of how to choose the intermediate probabilities $p_i(\phi)$ is presented in \cite{Neal1996}. In summary,  it is necessary that the probabilities $p_i(\phi)$ at the ``top" of the ladder in 
Fig.~\ref{fig:tempering_step_figure} {\it do not} have well separated modes and that the subsequent distribution probabilities ($p_i(\phi)$ and $p_{i+1}(\phi)$) be close enough  so the Monte Carlo chain along up and down the ladder in Eq.~\ref{eq:tempered_proposal} can gently guide the configuration from one mode to another.

In our implementation of tempered transitions, we modulate the probability distributions up the ladder by adjusting the amount of flow we subject the parameterization manifold to. It can be seen from Fig.~\ref{fig:srvsphi} that at $T_{\text{flow}}=0.0$, the action barriers on $\Man$ are mild while for large $T_{\text{flow}}$ the action barriers are high. Therefore, we choose $p_0(\phi)$ to be the probability distribution generated with a large enough $T_{\text{flow}}$ to tame the sign problem and we choose $p_n(\phi)$ to be the probability distribution generated when $T_{\text{flow}}=0.0$ where fields are mobile. There much latitude in how to choose the probability distributions between $p_0(\phi)$ and $p_n(\phi)$, but not all choices perform the same. We choose to interpolate between a $T_{\text{flow}}=T_{\text{max}}$ and $T_{\text{flow}}=0.0$ linearly. Whatever the optimal choice may be, it is clear that $p_{i}(\phi)$ and $p_{i+1}(\phi)$ should be similar in some sense for such a choice. This is because if \emph{all} $p_i(\phi)$ are the same then the tempered proposal is guaranteed to be accepted.

\section{Results}
\label{sec:results}
We now apply the method of tempered transitions on $0+1$ Thirring model at finite density. For non-zero chemical potential the determinant of the Dirac matrix is complex and the theory suffers from a sign problem. We discretize the Euclidean time direction using staggered fermions. The lattice partition function for this theory is 
\beq
Z=\left[\prod_{t=1}^{N} \int_{0}^{2\pi} {d \phi_t\over 2\pi}\right] \det D e^{-{1\over 2g^2} \sum_{t=1}^N (1-\cos \phi_t)}\equiv\left[\prod_{t=1}^{N} \int_{0}^{2\pi} {d \phi_t\over 2\pi}\right] e^{-S[\phi]}, \label{eq:Z-lattice}
\eeq
where the effective action and the Dirac matrix are explicitly given by 
\begin{eqnarray}
 S[\phi]&=&{1\over 2g^2} \sum_{t=1}^N (1-\cos \phi_t)-\log\det D \label{eq:S} \\
D_{t,t^\prime}&=&{1\over 2}\left(e^{\mu+i \phi_t}\delta_{t+1,t^\prime}-e^{-\mu-i \phi_{t^\prime}}\delta_{t-1,t^\prime}+e^{-\mu-i\phi_{t^\prime}}\delta_{t,1}\delta_{t^\prime, N} -e^{-\mu-i\phi_{t}}\delta_{t,N}\delta_{t^\prime, 1}\right)+m\,\delta_{t,t^\prime}\,. \label{eq:K}
\end{eqnarray}
Here $N$ is an even number equal to to the number of lattice sites and all dimensionful quantities are measured in units of the lattice spacing $a$ which we set to one. This discretized model is exactly solvable~\cite{Pawlowski:2013pje};  an observable of interest  is the chiral condensate, which is given by:
\beq
\langle \bar\psi\psi \rangle = {1\over N}{\partial \over \partial m} \log Z= {(1+m^2)^{-1/2}I_0^N({1 \over 2g^2})\sinh(N\sinh^{-1}(m)) \over I_1^N({1 \over 2g^2})\cosh(N\mu)+I_0^N(\alpha) \cosh(N \sinh^{-1}(m)) }
\eeq
where $I_n(z)$ denotes the modified Bessel function of the first kind of order $n$. We will use this exact solution to compare against. 

\begin{figure}[t]\captionsetup{singlelinecheck = false, justification=raggedright}
\includegraphics[scale=0.35]{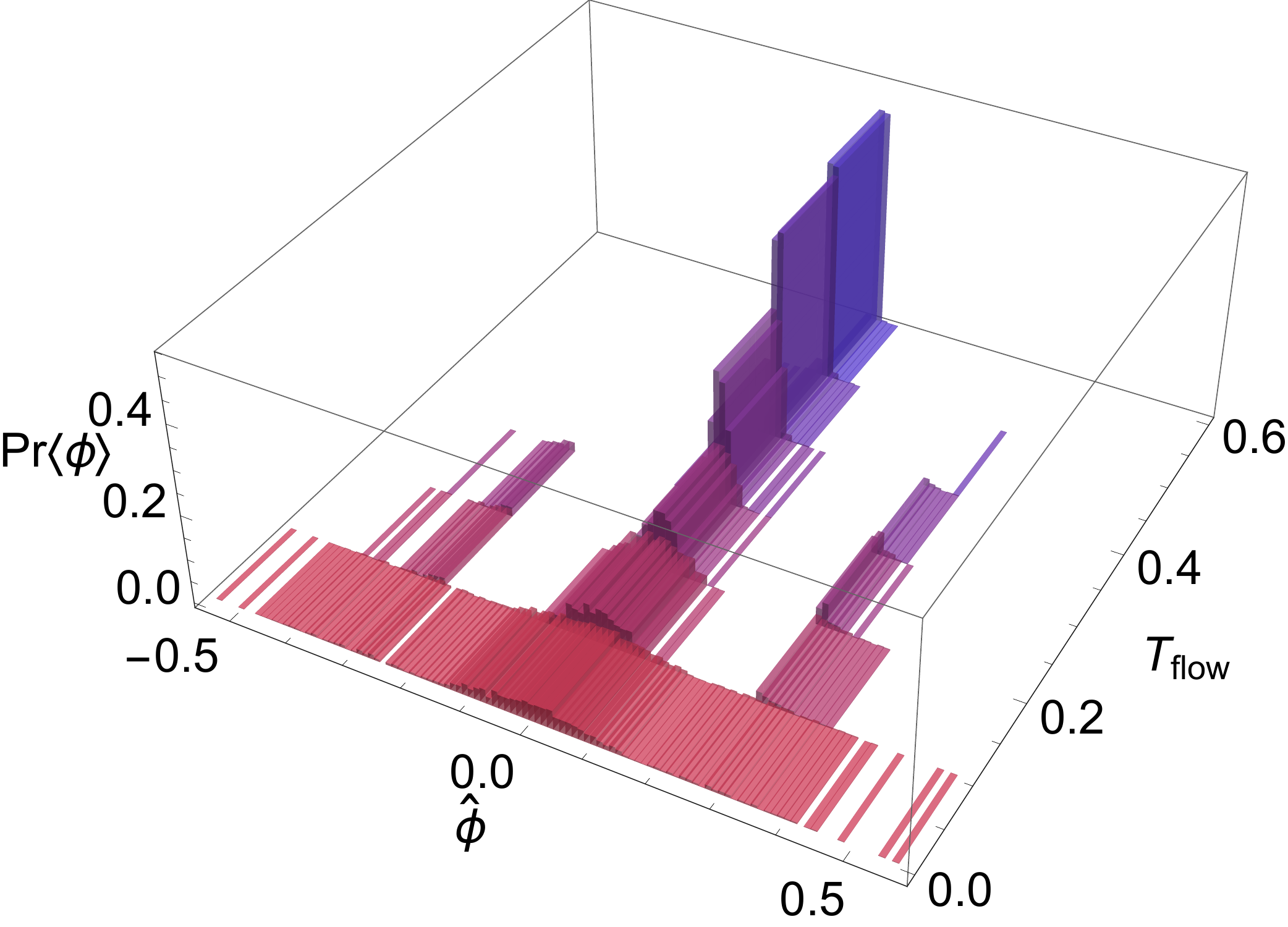}
\caption{
A demonstration of the trapping process. We show histograms of the time average  of the field $\hat{\phi} = {1 \over N}\sum_{k=1}^N{\phi_k}$ for simulations performed in a range of six flow times ($T_{\text{flow}}=0.0,0.1,...,0.5$). The widest histogram corresponds to a Monte Carlo run at zero flow on the tangent plane of the global minimal thimble and the narrowest histogram corresponds to a Monte Carlo run at $T_{\text{flow}}=0.5$. Beyond $T_{\text{flow}}=0.3$, the Monte Carlo is unable to tunnel through to neighboring thimbles and misses their significant contributions. We have demonstrated that at the $T_{\text{flow}}=0.5$ this trapping is for all practical purposes indefinite. We will use the  value $T_{\text{flow}}=0.5$ to illustrate how the tempered transition algorithm allows for the proper sampling of the space even in this case. }
\label{fig:trapping}
\end{figure}

\begin{figure*}[t]\captionsetup{singlelinecheck = false, justification=raggedright}
\includegraphics[scale=0.37]{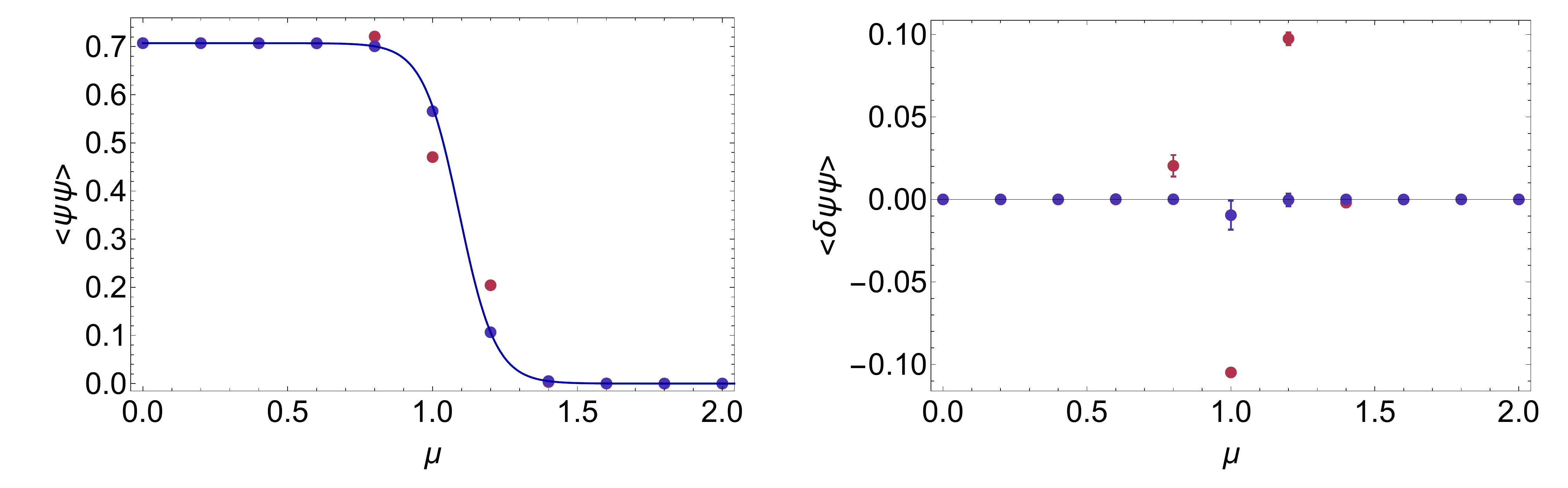}
\caption{In the left panel we have the numerical computation of the chiral condensate $\langle \bar{\psi} \psi \rangle$ at $T_{\text{flow}}=0.0$ (shown in blue) and $T_{\text{flow}}=0.5$ (shown in red) for the parameters $N_t=16$, $g^2 = {1 \over 6}$ and $m=1.0$. The exact result is shown as the solid line. In the right panel we plot the difference between the exact result and the numerical result with the same color coding. The discrepancy between the exact result and the numerical result is easily visible for chemical potentials near the transition.}
\label{fig:chi_cond}
\end{figure*}

As an example we take an $N=16$ lattice with $m=1.0$ and $g^2 = {1 \over 6}$ over a range of chemical potentials. First, we reduce the sign problem for these parameters by shifting the domain of integration from the real hyperplane to the tangent plane of the thimble fixed to the global minimum  of the action \cite{Alexandru:2015sua} where the phase fluctuations are smaller than on the real plane. The remaining phase fluctuations can be tamed by flowing the tangent plane. 
Calculations performed with flow times in the range $0.2 \alt T_{\text{flow}}\alt 0.4$ shows that at least three thimbles contribute significantly (see Fig.~\ref{fig:trapping}). For larger values of flow time, calculations performed with the Metropolis algorithm remain trapped in the region with $\langle \phi\rangle\approx 0$ for  a very large number of steps (we have followed the Monte Carlo chain up to  $10 \times 10^6$  steps without seeing a transition to other thimbles).
 For some of  the parameters explored, in particular when the chiral condensate drops quickly around the mass of the fermion $\psi$, computing observables when trapped to the global minimal thimble yields statistically incorrect results, see Fig.~\ref{fig:chi_cond}. The need to integrate over several thimbles at points of sharp variation in thermodynamic functions has been stressed before in fermionic models \cite{Tanizaki:2016cou}. 
As the discrepancy between the trapped numerical result and the exact result is largest at $\mu=1.0$, we restrict our further analysis to this value of the chemical potential. 
We apply  the method of tempered transitions using a ladder with 1600 steps in each direction. At the bottom of the ladder the flow is 0.5 where the fields are trapped to a single thimble, and at the top the flow is 0.0 were the Monte Carlo chain explores all thimbles. We interpolate linearly in the flow as sub-transitions are applied. At each step of the ladder, we use a transition operator $T_k(\phi,\phi')$ which applies 10 standard metropolis proposals at the flow time $t_k$ to the configuration $\phi$.  The reason for using such an operator is to allow the newly relaxed distribution to equilibrate before relaxing it further. We find that this procedure reduces the ``free energy" cost of a tempered transition and increases the likelihood of a tempered transition to be accepted. 

There are two time scales in the equilibration of the Monte Carlo chain: a fast one for the equilibration within one mode and a slow one for the equilibration between modes. These separated time scales exist for the simple reason that only small proposals are required to explore an individual thimble while a large proposal is required to transition. This is a generic property of theories with multiple thimbles. In light of this observation, we construct our Monte Carlo as follows: between two tempered transitions we perform 1000 standard metropolis steps. This allows all relevant thimbles to be explored and the  space within each thimble to be explored. Moreover such a division is necessary because tempered transitions are expensive compared to normal transitions. In spite of this cost, tempered proposals induce transitions at a substantially higher rate than standard Metropolis proposals. In this study, a tempered transition is composed of $10\times3200 = 3.2\times 10^4$ individual metropolis steps. We find that roughly every $10^{\text{th}}$ proposal between the main thimble and the shoulder thimbles is accepted, so it takes the comuptational effort of $\sim 3\times10^5$ metropolis steps to induce a transition. Therefore tempered proposals are at least $1-2$ orders of magnitude more efficient than standard metropolis proposals in this study.


\begin{table}[h!]\captionsetup{singlelinecheck = false, justification=raggedright}
\centering
\begin{tabular}{| c  |l |}
\hline
 {Method} &{$\langle \bar{\psi}\psi \rangle$} \\
 \hline
 Exact & 0.575 \\
 \hline
 with tempered transition & 0.602(20) \\
 \hline
 w/o tempered transitions & 0.470(02) \\
 \hline
\end{tabular}
\caption{Results for the chiral condensate at $\mu=1.0$ for the exact solution, a Monte Carlo trapped to the global minimum thimble and a Monte Carlo utilizing tempered transitions. }
\label{tab:theresult}
\end{table}

\begin{figure*}[t]\captionsetup{singlelinecheck = false, justification=raggedright}
\includegraphics[scale=0.5]{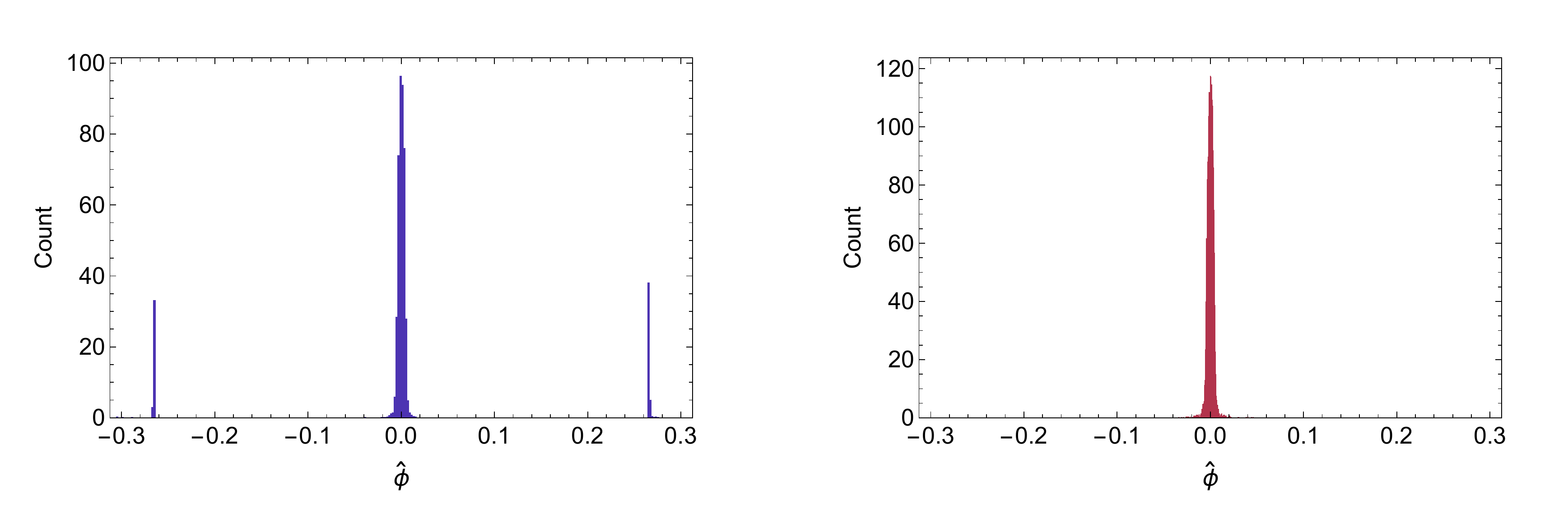}
\caption{Here we show the distribution over the course of a simulation of the time average in the tempered $($left$)$ and non-tempered $($right$)$ case. The two Monte Carlos are of equal length and sample from the same sharply peaked probability distribution with isolated regions of support. The simulation without tempered steps misses the two side peaks.}
\label{fig:temper}
\end{figure*}

\begin{figure*}[t]\captionsetup{singlelinecheck = false, justification=raggedright}
\includegraphics[scale=0.5]{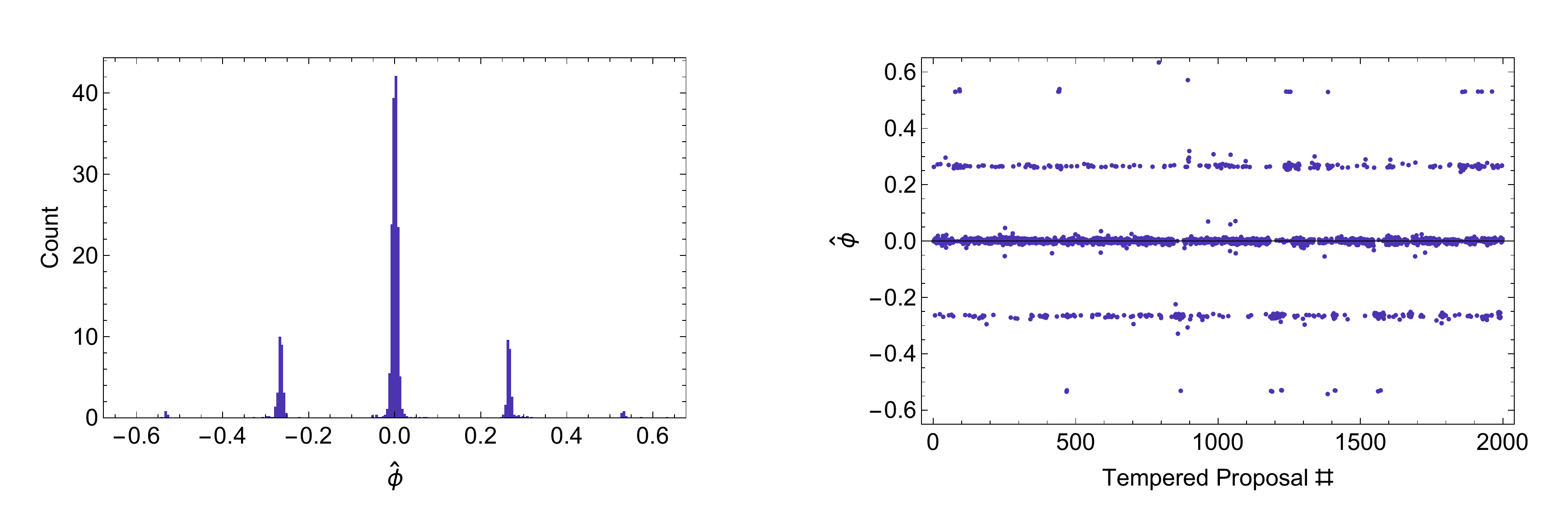}
\caption{Proposals made by tempered steps (projected on the direction) as a function of Monte Carlo step (right panel) and its histogram (left panel). Notice that proposals are made mainly to five different regions, corresponding to the five dominating thimbles. The two next-to-nearest thimbles to the central one are not accepted, in accordance to their small statistical weight.}
\label{fig:five_thimble_temper}
\end{figure*}

The results of a Monte Carlo using 2000 tempered transitions is shown in Fig.~\ref{fig:temper}. We find that the combination of tempered transitions with 1600 ladder steps and regular Metropolis steps is sufficient to transition regularly between thimbles. In addition, we find that the inclusion of the neighboring thimbles reproduces the exact result up to statistical errors as can be seen in Table \ref{tab:theresult}. Roughly $10\%$ of tempered transitions from the central thimble to the shoulder thimbles were accepted. Note however that this low acceptance rate is not due to poor proposals but because the shoulder thimbles carry roughly $10\%$ of the weight of the path integral. In fact, the tempered transitions make proposals to the next-to-nearest shoulder thimbles frequently. This can be seen in Fig.~\ref{fig:five_thimble_temper} where the tempered proposals are plotted as a function of Monte Carlo step. These transitions to the next-to-nearest shoulder thimbles are, however,  not accepted because these thimbles contribute very little to the path integral. It is expected, then, that in a problem where many thimbles carry substantial weight, tempered transitions may provide a natural means to explore many thimbles without any \emph{a priori} knowledge of their location. 

\section{Conclusion}
\label{sec:conclusion}

The holomorphic gradient flow approach to the computation of path integrals with a sign problem frequently leads to multimodal probability distributions. Each of the modes are related to a thimble contributing to the integral, as described by Picard-Lefschetz theory. This poses a challenge to numerical computations as Monte Carlo chains tend to get ``stuck" in one of the modes for exponentially long times. 
We applied the method of tempered transitions to this problem. We take the tempering parameter, which controls the steepness of the probability landscape in each ladder of the tempering process,  to be the flow time \cite{Fukuma:2017fjq} by which the real space is transported by the holomorphic gradient flow. We demonstrated in a simple example that this procedure is feasible and that it allows for the proper sampling of the field space even in circumstances where a simpler Metropolis algorithm fails. It was found that a combination of tempered steps interspersed with regular Metropolis steps was the most efficient choice. 

The method is not without drawbacks. The main one is the need, even in the simple model considered here, of a very large number of ladders steps during a tempered proposal and the associated large computational cost. This is the main difficulty that limited us in this paper to fairly small toy models. Our experience in scaling up the number of  degrees of freedom is that the number of ladder steps required scales roughly linearly with the number of degrees of freedom. By itself this is not such a steep scaling but it should be kept in mind that other, steeper increases in cost are caused by the computation of the jacobian, increase in required flow time and, in models with more spacetime dimensions, increase on the size of the Dirac matrix (in the $0+1$ dimensional model discussed here the determinant has a closed form). This is not to say that significant improvements are not possible by adjusting some parameters in our simulations. For instance, there is tremendous latitude in choosing the values of the intermediate flows. Most applications of the tempered transitions method use the temperature as the parameter changing along the tempered ladder, leading to an exponential flattening of the probabilities distributions \cite{Neal1996}. A similar exponential flattening of probabilities can be obtained by choosing the intermediate flows to be equally spaced along the ladder. It could well be that a different choice is significantly better.

As we finished the present paper a study of the same model we discuss using a different tempering method appeared \cite{Fukuma:2017fjq}. A direct comparison  of the efficacy of the two methods is hindered by some small differences between  the two calculations that are unrelated to the tempering method. For instance, in \cite{Fukuma:2017fjq}, the manifolds of integration are obtained by flowing the real manifold, not the main tangent manifold as is done here. It is clear, however, that either method will be extremely costly when applied to realistic field theories.


\acknowledgements

A.A. is supported in part by the National Science Foundation CAREER grant PHY-1151648 and by U.S. DOE Grant No. DE-FG02-95ER40907.
A.A. gratefully acknowledges the hospitality of the Physics Department at the University of Maryland where part of this work was carried out.

G.B. is supported by the U.S. Department of Energy under Contract No. DE-FG02-01ER41195. 

P.B. and N.C.W. are supported by U.S. Department of Energy under Contract No. DE-FG02-93ER-40762.

\medskip

\bibliographystyle{JHEP}
\bibliography{temperedbib}

\end{document}